
\def\<{\langle} \def\>{\rangle} \def\br{\bf\rm} \def\cl{\centerline}
\def\it{\tenit}   
\def\half{{\scriptstyle{ 1\over 2}}}  \def\Re{{\rm Re}}
\def\I{{\rm I}}   \def\R{{\rm R}}
\def\M{{\br M}}  \def\d{{\rm d}} \def\s{{\rm s}}
  
 \def\hs1{\hskip1mm}
\def\h10{\hskip10mm} \def\dag{^\dagger}

\magnification1300

\parskip 2mm plus 1mm \parindent=0pt

\def\TP{T_{\rm P\ell}} \def\tb{\overline t}
\def\Ka{K\'arolyh\'azy}


 {\it 94icppapers/95stpsd4}

\vskip15mm \bf

\centerline{Quantum space-time fluctuations and primary state diffusion}
{\vskip2mm}

\centerline {by}

\vskip2mm

\centerline {Ian C Percival}
\vskip2mm
\centerline {Department of Physics}
\centerline {Queen Mary and Westfield College, University of
London}
\centerline {Mile End Road, London E1 4NS, England}

\vskip1mm
\cl{and}
\vskip1mm
\cl{Albert-Ludwigs Universit\"at - Physik}
\cl{Herman-Herder Strasse 3}
\cl{D-79109 Freiburg i Br., Germany}

\vskip15mm \centerline{Abstract}

\rm

Nondifferentiable fluctuations in space-time  on a Planck
scale introduce stochastic terms into the equations for quantum
states, resulting in a proposed new foundation for an existing alternative
quantum theory, primary state diffusion (PSD).  Planck-scale
stochastic space-time structure results in quantum fluctuations, whilst
larger-scale curvature is responsible for gravitational forces. The
gravitational field and the quantum fluctuation field are the same,
differing only in scale. The quantum mechanics of small systems,
classical mechanics of large systems and the physics of quantum
experiments are all derived dynamically, without any prior division into
classical and quantum domains, and without any measurement
hypothesis.  Unlike the earlier derivation of PSD, the new derivation,
based on a stochastic space-time differential geometry, has essentially no
free parameters. However many features of this
structure remain to be determined. The theory is falsifiable in the
laboratory, and critical matter interferometry experiments to distinguish
it from ordinary quantum mechanics may be feasible within the next
decade.

\vfill

95 July 17,  QMW TH 95-20  \hfill  Revised version, resubmitted to Proc.
Roy. Soc. A  \eject

\vskip10mm

{\bf 1.  Introduction}

{\bf 2. Classical diffusion and quantum state diffusion}

{\bf 3. Fluctuating space-time and equivalence principles}

{\bf 4. Theories of quantum decoherence from gravity}

{\bf 5. Sharp quantum theories and experiment}

\vskip10mm

{\bf 1.  Introduction}

This paper is intended to be read in conjunction with the original paper on
primary state diffusion (Percival, 1994b:  PSD1), whose contents are
briefly summarised in the next four paragraphs.

PSD1 describes an alternative quantum theory  (PSD) developed from
quantum state diffusion theory (Gisin \& Percival, 1992,1993a,b; Percival
1994a), in which both quantum and classical behaviour are derived from
the same dynamics. Decoherence,  localization (reduction) and the
interaction of `quantum' system with `classical' apparatus are derived
from the dynamics, and no independent measurement hypothesis or
interpretation is required. In PSD, unlike the earlier quantum state
diffusion theory, the diffusion is  independent of any division between
system and environment or observed system and observer.  Primary state
diffusion, like many similar theories referred to in PSD1 and in sections 4
and 5 of this paper, provides a common dynamical basis for classical and
quantum systems and for their interaction, for example in the laboratory
measurement of the state of a quantum system.

The distinguishing feature of PSD1 is that the diffusion of quantum states,
which dominates on very small times scales, is primary, and the usual
Schr\"odinger evolution of the state vector is derived from it.  In the
nonrelativistic theory, the stochastic differential equation for the state
vector incorporates a time-dependent fluctuation and results in energy
localization.  This energy localization produces decoherence, but does not
produce the  phase space (including position)  localization needed for the
dynamics of classical systems.  But special relativity requires that the
quantum fluctuations should depend on space as well as time.  This space
dependence of the fluctuations gives the position localization.

The theory of PSD1 contains an undetermined time constant $\tau_0$,
whose value was conjectured to be approximately the Planck time:
$$
 \TP = (\hbar G c^{-5})^{1/2} \approx 5\times 10^{-44}\s.
\eqno(1.1)$$
Order of magnitude estimates then show localization of wave packets of
caesium atoms over a distance of 15 cm, probably within the
range of feasible matter interferometry experiments within the next
decade.

The theory of PSD1 was derived without any reference to the
structure of space-time.

In this paper it is shown that PSD leads naturally to  stochastic
differential relations for time, and thus, by a necessary extension, for
space-time.  This leads directly to a derivation of the previously
conjectured value of $\tau_0$.

In the new formulation of PSD of this paper, the  foundations
are changed and the theory is derived from stochastic
fluctuations of space-time on the scale of the Planck length and time and
below. On these scales the time-time component of the transformation
from a Minkowski universe to our universe resembles the transformation
from the time variable to the space variable of a Brownian motion.
  This transformation is not differentiable and requires the
Ito differential calculus to replace ordinary differential calculus.

By invariance the same may be said of the other components of the
transformation, and of the intrinsic geometry of space-time, which
is then a stochastic differential goemetry. But these remain to be
determined.

The representation of quantum fluctuations by
a change in the differential structure of space-time differs from
proposals using a quantized gravitational field (e.g. Grishchuk, Haus \&
Bergman, 1992) and from proposals to retain a smooth
differential structure and rely on wormholes (Ellis, Mohanty \&
Nanopoulos, 1989).

In the original PSD equation (2.12), the
fluctuations $\d\xi$ are coupled to the quantum system through the
hamiltonian; that is, relativistically, they are coupled to matter through
the mass. But the gravitational field is also coupled to  matter through the
mass.  This suggests that quantum fluctuations and gravity might be
different manifestations of the same field.  In general relativity, gravity
comes from the large scale structure of space-time, so we should expect
the quantum fluctuations to come from the small scale structure of
space-time.

This results in a semiclassical theory in which
matter is treated quantally, and space-time is represented classically.
The probabilistic quantum properties of matter follow from the
stochastic space-time fluctuations, which drive the quantum state
through the resultant diffusion term which is added to the Schr\"odinger
equation.  With this hypothesis, we obtain a space-time primary state
diffusion theory in which the results of the original PSD theory are
rederived, with the one difference that $\tau_0$ is shown to be close to
the Planck time $\TP$.   The only important constants to appear in the
theory are the velocity of light, the gravitational constant $G$ and the
Planck constant $\hbar$ for quantum mechanics.

However, this space-time PSD theory is far from complete, since it is
worked out in detail only for the internal quantum dynamics of those
systems for which all but the time-time component of space-time
transformations can be neglected, and the effects of the spatial
dependence of these components can also be neglected (Pauli 1967 p150,
Weinberg 1972 p78).  For a more general theory, nonrelativistic quantum
state diffusion theory could be used as guide.  For nonlinear systems, the
fluctuations in the classical phase space trajectories of means of
dynamical variables can be obtained.  These fluctuations do not
satisfy Hamilton's equations, so we should not expect the space-time
fluctuations of PSD to satisfy Einstein's equations.  However, since the
large-scale structure of space-time is derived from the Planck-scale
structure, the approximate validity of Einstein's equations in the large
will put a constraint on the small-scale theory, just as the Schr\"odinger
equation in the large put a constraint on the form of the quantum state
diffusion in PSD1.

The development of modern alternative quantum theories has been
hampered because they have little or no contact with
experiment.  Either their predictions agree with those of ordinary
quantum theory, or they contain parameters whose values can be shifted
to ensure that agreement, for any experiments in the foreseeable future.
This is true of the original PSD theory of PSD1.  By contrast space-time
primary state diffusion theory has essentially no free parameters
and could be falsified by matter interferometry experiments within the
next decade (PSD1, Section 9).

Section 2 contains a short account of Langevin-It\^o theory
(see Gardiner, 1985, 1991),  quantum state diffusion  and PSD, but the
reader who needs to study the details, particularly the property of
localization and the resultant derivation of classical mechanics from
quantum mechanics with state diffusion, should consult the references
in that section.

Section 3 is the key section, since it introduces fluctuating space-time
and a related equivalence principle, which replaces the equivalence
principle of general relativity (GR) on scales less than or equal to the
Planck scale.  In this section it is shown that the resultant equations for
a quantum state are the PSD equations with $\tau_0\approx\TP$.

Sections 4 and 5 discuss alternative quantum theories and
experimental tests. Section 4  reviews gravitational theories of quantum
decoherence, including that of \Ka (1966), which most resembles
space-time PSD.  Section 5 is a guide to other current sharp formulations
of quantum mechanics, as defined by Bell (1987, p171) to mean a
formulation that provides a uniform description of the micro and macro
worlds. The emphasis is on critical experiments needed to distinguish such
sharp theories from ordinary quantum mechanics, including the critical
matter interferometry experiments for space-time PSD.
 \vfill\eject

{\bf 2. Classical diffusion and quantum state diffusion}

It\^o has developed a very elegant theory and differential calculus for
stochastic processes that satisfy Langevin equations,  described
for physicists in the books of Gardiner.  For example when a
particle with displacement $s$ diffuses in one dimension with no drift,
as in Brownian motion, the mean of the distance traversed is zero and the
mean square  is proportional to the time.  So for small displacements $\d
s$,  $$
\M\d s(t) = 0, \h10 \M[\d s(t)]^2 = a^2\d t,
\eqno(2.1)$$
where $a$ is a constant and $\M$ represents a mean over the
ensemble of all possible processes.  For more complicated stochastic
processes,  it is convenient to introduce a standard normalized
stochastic differential $\d w$, with zero mean and mean square $\it
equal$ to the time, so that
$$
\M\d w = 0, \h10 \M(\d w)^2 = \d t.
\eqno(2.2)$$
For the example of one-dimensional Brownian motion with drift velocity
$v$ and diffusion constant $a^2$, the Langevin-It\^o differential equation
is written as
$$
\d s = v\d t + a\d w,
\eqno(2.3)$$
where the coefficient of $\d t$ gives the drift and the coefficient of $\d
w$ the diffusive part of the motion.

For applications to quantum mechanics it is convenient
to consider two-dimensional isotropic diffusion as one-dimensional
complex diffusion in the  complex $z$-plane.  The standard
normalized complex stochastic differential is
$$
\d\xi = \d\xi_\R + i \d\xi_\I,
\eqno(2.4)$$
where the real and imaginary parts satisfy
$$
\M\d\xi_\I = \M\d\xi_\R =  0, \h10 \M\d\xi_\R^2 = \M\d\xi_\I^2 \
=  \d t/2.
\eqno(2.5)$$
More convenient is the equivalent complex form of the conditions:
$$
\M\d\xi = 0, \h10 \M(\d\xi)^2 = 0,\h10\M|\d\xi|^2 = \d t,
\eqno(2.6)$$
where the statistics of the motion is unaffected if $\d\xi$ is multiplied
by a phase factor of unit norm.

The Langevin-It\^o  equation of a particle with drift and diffusion at
position $z(t)$ in the complex plane is
$$
\d z = v \d t + a \d\xi.
\eqno(2.7)$$
 Both $v$ and $a$ can be complex, but the physics
is unaffected by the (gauge) transformation for which $a$ is multiplied
by a phase factor.

Also for more general differential stochastic processes, the
drift is the term containing $\d t$ and the diffusion the term
containing $\d\xi$. For sufficiently small times, the diffusion, which is
proportional to $(\d t)^{1/2}$, dominates the drift, which is proportional
to $\d t$,  and when using the It\^o calculus great care must be taken
to ensure that quadratic terms with a factor $|\d\xi|^2$ are
included. This problem is overcome in the alternative but equivalent
Stratonovich calculus (Gardiner 1985), at the price of a slightly more
difficult interpretation of the results, but the It\^o form is used here.

Quantum state diffusion represents the stochastic evolution of individual
systems of an ensemble in states $|\psi\>$, when the  ensemble as a
whole is usually represented by a density operator $\rho$.  In
quantum state diffusion theory, unlike the usual density
operator theory, care must be taken to distinguish the quantum
expectation  $\<\dots\>$ for an individual system from the mean $\M$
over the ensemble.

The simplest master equation for $\rho$, with no Hamiltonian evolution,
is
$$
 \dot\rho = L\rho L\dag - \half L\dag L\rho - \half\rho L\dag L,
\eqno(2.8)
$$
where $L$ is a Lindblad operator which may or may not be
self-adjoint.

The corresponding state diffusion equation is
$$
 |\d\psi\> = \big(\<L^{\dagger}\> L
-\half L^{\dagger} L - \half\<L^{\dagger}\>
\<L\>\big) |\psi\>\d t + \big(L -\<L\>\big) |\psi\>\d\xi,
\eqno(2.9)$$
where $\<L\> = \<\psi|L|\psi\>$.  Gisin and Percival (1992) derive the state
diffusion equation (2.9) from the master equation (2.8). The derivation
requires $|\psi\>$ to be normalized, and also requires the state diffusion
equations, like the master equation, to be invariant under a unitary gauge
transformation $L' = u L$ in operator space. With these constraints, the
set of equivalent state diffusion equations corresponding to a given
master equation is  unique.

For sufficiently short times the QSD evolution is dominated by the
diffusion term, which represents one-dimensional complex diffusion
on the sphere of normalized quantum states.

For a nonrelativistic system with hamiltonian $H$, the  PSD master
equation and primary state diffusion equation have the forms (2.8, 2.9)
with  $$
L = \tau_0^{1/2}{ H \over\hbar} +  i\tau_0^{-1/2}I,
\eqno(2.10)$$
where $I$ is the identity operator, but it is more convenient to write
them explicitly in terms of the Hamiltonian as the transformed equations
(PSD1)
$$
\hbar\dot\rho = -i[H,\rho] + \tau_0(H\rho H -\half H^2\rho -\half\rho
H^2)
\eqno(2.11)$$
$$
\hbar|\d\psi\> = (-iH_\Delta \d t- \half\tau_0 H_\Delta^2\d t
                + \tau_0^\half H_\Delta \d\xi)|\psi\>,
\eqno(2.12)$$
where
$$
H_\Delta = H - \<\psi| H |\psi\>.
\eqno(2.13)$$
Note that in  PSD1 the transformed equation
corresponding to (2.12) has a factor $\half$ missing.

The additional fluctuation term destroys the Fourier relation between
energy representation and time representation.  In addition to the usual
exponential of the Fourier transform there is a fluctuating term.  In a
relativistic theory this must also be true of momentum and position
representation.  So for high frequencies and large wave numbers, the
usual simple relations between these representations break down.
\vfill\eject
{\bf 3. Fluctuating space-time}

According to the equivalence principle of general relativity (GR),
around every space-time point of our universe there is a sufficiently
small region with a smooth transformation relating it to a flat Minkowski
universe.

It is commonly agreed that in our universe,  space-time has quantum
fluctuations on the Planck scale, and that by quantum indeterminacy
these fluctuations become relatively larger on  smaller scales.  It
follows that they cannot be differentiable.  Accordingly there are no
locally inertial Lorentz frames in our universe, and no small region
around any point with a smooth transformation relating it to a Minkowski
universe.   As we go down in scale into the Planck domain,  space-time
become less flat, not more so.  The GR equivalence principle applies only
on scales significantly larger than the Planck scale.

Space-time PSD is based on a stochastic differential structure
and a fluctuating space-time metric.   These are quantum fluctuations, but
they are expressed in terms of non-differentiable relations between
classical space-times.  All other quantum fluctuations are consequences of
these space-time fluctuations.   Space-time PSD  is therefore made of two
parts: the properties of a Minkowski universe and its Lorentz frames, and
the properties of our universe that are obtained by local nondifferentiable
space-time transformations between the Minkowski universe and ours.

Just as in GR,  our universe can be locally related to a Minkowski
universe by a space-time transformation, even down to the Planck scale
and below.  But whereas on a macro scale, the space-time of our universe
and the transformations appear to be smooth and differentiable, so on the
Planck scale the space-time and the transformations are fluctuating and
nondifferentiable. In the present semiclassical theory they resemble
Brownian motion and satisfy Langevin-It\^o stochastic differential
relations.

In a Lorentz frame in the Minkowski universe with time $\tb$, elementary
quantum systems at low velocity are represented by states that satisfy
the Schr\"odinger equation,
$$
\hbar|\d\psi\> = -i H\d \tb |\psi\>
\eqno(3.1)$$
and the more subtle systems satisfy the flat space-time laws of quantum
field theory.  There is no classical dynamics because there is no
quantum state diffusion and no localization.  In PSD, as in other sharp
theories, there is only one world, which includes both quantum and
classical dynamics, but for PSD the existence of the latter depends on the
localization in phase space that comes from the quantum state diffusion.

To help us in the journey from the Minkowski universe to our universe it
is convenient to introduce an intermediate stage.  This is the uniformly
fluctuating universe or UFU.  The local transformation from the
Minkowski universe to the UFU is stochastic, nondifferentiable and
nonuniform in space-time, but it has uniform statistical properties in
space-time. This represents quantum fluctuations. The local transformation
from the UFU to our universe is deterministic, smooth and nonuniform in
space-time.  This represents large-scale gravity.

The fluctuation equivalence principle that replaces the usual GR
equivalence principle says that around every space-time point of our
universe there is a sufficiently small region with a smooth
transformation relating it to the UFU. This sufficiently small region is
normally very large compared to the Planck scale, but the transformation
applies on all scales, and is the equivalent, in the new theory, of the
smooth transformations of GR.

We look at the new quantum physics coming from the
nondifferentiable transformations between the flat Minkowski universe
and the UFU.  Since the theory is not complete, we do not have the
statistics of all components of this transformation, nor of its
space dependence, which would then describe the statistical metric
structure of the UFU in full.  We consider only a local time-time
component of the transformation from Minkowski space-time with time
$\tb$ to UFU space-time with time $t$.

Suppose that on far sub-Planck scales, much smaller than the Planck
scale, the transformation is dominated by the fluctuation term,  that on
far super-Planck scales it is dominated by the identity transformation,
and that on the Planck scale the two effects are about equal. The
time-time component of the fluctuating transformation is then
$$
 \d\tb = \d t + \tau_1^\half \d\xi,
\eqno(3.2)$$
where $\tau_1$ is a universal time constant.  The simplest assumption
is that this is the Planck time $\TP$, but a small numerical
factor $C$ of order unity, like $2\pi$, cannot be excluded, so
$$
\tau_1 = C \TP.
\eqno(3.3)$$

In classical dynamics a real time-time transformation has no local
 physical effect other than a change in the time coordinate that results
in a time dilation or contraction with respect to other frames, as in
the usual time dilation of relativity.  But for complex time
transformations  of quantum systems this is no longer so.  They produce
nonphysical changes in the norm of the state vector, which must be
removed.    This is done through a set of conditions on the state vector of
space-time PSD which is must be satisfied in all frames.  These
conditions are included in the principles of space-time PSD, or STP,
which must define the time evolution of the state uniquely.  For the UFU
the principles are

STP1.  The state vector $|\psi\>$  satisfies a nonlinear Langevin-It\^o
diffusion equation,  where the right side is an operator on
$|\psi\>$.

STP2.  On far sub-Planck scales the  operator is given by the
transformation from the flat Minkowski universe, except for an additional
scalar times the unit operator, which is needed to preserve
normalization.

STP3.  On far super-Planck scales the skew-adjoint part of the
operator is given by the transformation from the Minkowski universe.
This is the Schr\"odinger operator.

STP4.  The norm of $|\psi\>$ is preserved.

The principles for our universe are obtained from the STP principles
for the UFU and the new equivalence principle.

{}From STP1-3, using the Schr\"odinger equation (3.1) in Minkowski space,
and the transformation (3.2), the UFU state vector satisfies an equation of
the form
$$
\hbar|\d\psi\> = (-i H\d t + R(\psi)\d t + \tau_1^{1/2} H\d\xi -
s(\psi)\d\xi)|\psi\>,
\eqno(3.4)$$
where $R(\psi)$ is a self-adjoint operator.

It follows from STP4 that
$$\eqalign{
0 = \d\<\psi|\psi\> = 2\Re\<\psi| -iH &+ R(\psi) |\psi\>\d t
                          + 2\Re(\<\psi| \tau_1^{1/2} H-sI |\psi\>\d\xi) \cr
    &+\<\psi|(\tau_1^{1/2} H-sI)\dag (\tau_1^{1/2} H-sI )|\psi\>\d t
}\eqno(3.5)$$
and since $\d\xi$ can vary, its coefficient must be zero,
so $s=\tau_1^{1/2}\<\psi| H |\psi\>$ and
$$
R(\psi) = -\half\tau_1 H_\Delta^2,
\eqno(3.6)$$
with $H_\Delta$ defined by (2.13),
so the state vector in the uniformly fluctuating universe satisfies
the primary state diffusion equation (2.12) with
$$
\tau_0 = \tau_1 = C\TP,
\eqno(3.7)$$
as required.

The Lindblad-Gisin condition  of  PSD1, requiring the
derived  density operator equation to be of Lindblad (1976) form,  is not
needed here as one of the principles of space-time PSD.  It is satisfied
automatically.

For a restricted set of problems, space-time PSD unites space-time
structure and the foundations of quantum theory.  It does so by
abandoning one of the basic principles of each.  The principle of
equivalence of general relativity relating our universe to a Minkowski
universe is replaced by a different principle relating our universe to a
uniformly fluctuating universe.  The Fourier relations between
energy-momentum and space-time representations in quantum mechanics
are also lost.  So for these small scales the theory of space-time and
quanta must be built again on the new foundations.
\vfill\eject
{\bf 4. Theories of quantum decoherence from gravity}

In Feynman's lectures on gravitation  (Feynman et al 1961-2, p12), he
said

`The extreme weakness of quantum gravitational effects now poses some
philosophical problems; maybe nature is trying to tell us something here,
maybe we should not try to quantize gravity. \dots.  It is
still possible that quantum theory does not absolutely guarantee that
gravity {\it has} to be quantized. \dots In this spirit I would like to
suggest that it is possible that quantum mechanics fails at large
distances and for large objects.  Now, mind you, I do not say that I think
that quantum mechanics {\it does} fail at large distances, I only say that
it is not inconsistent with what we know. If this failure is connected
with gravity, we might speculatively expect this to happen for masses
such that $GM^2/\hbar c =1$, or $M$ near $10^{-5}$ g'

This is the Planck mass.  He continues later:

`If there was some mechanism by which the phase evolution had a little
bit of smearing in it, so it was not absolutely precise, then our
amplitudes would become probabilities for very complex objects.  But
surely, if the phases did have this built in smearing, there might be some
consequences to be associated with this smearing.  If one such
consequence were to be the existence of gravitation itself, then there
would be no quantum theory of gravitation, which would be a terrifying
idea for the rest of these lectures.'

This theme was taken up by \Ka\hs1  (1966; \Ka\hs1  Frenkel and Luk\'acs,
1986),  who connected stochastic reduction of the wave
function with gravity through an imprecision in the
space-time structure, just as it is here,  but the imprecision comes
from the presence of large masses, so the resultant decoherence is much
weaker.  There the matter rested for many years.

Recently Di\'osi and Luk\'acs (1993a, 1993b) showed that the Fourier
expansion of the small scale fluctuations in \Ka's theory leads to
unacceptably large fluctuations in energy density. The same objection
applies to space-time PSD.  However, the objection appears to be based on
the Fourier relation between space-time and energy-momentum, which is
no longer valid in PSD or related theories. If it were, then the white
noise of the fluctuations $\d\xi$ would produce infinite energies even in
the nonrelativistic quantum state diffusion theory (Gisin and
Percival, 1992), and it does not. Nevertheless the objection shows how
important it is to find a replacement for these Fourier relations.

Penrose (1986) based his gravitational theory of quantum
decoherence on the entropy relations of Hawking (1982), and the need for
gravitational entropy.  A wave function that is sufficiently spread out in
space, and gets coupled to a larger system which produces significant
Weyl curvature, represents a significant rise in the gravitational entropy,
and then reduction takes place.

Di\'osi (1987, 1989, 1992) emphasized that both gravity and
quantum mechanics must be changed because neither is valid on
all scales, and suggested a classical fluctuating gravitational field,
whose fluctuations are given by Heisenberg's indeterminacy principle.
Ghirardi, Grassi \& Rimini (1989) showed that Di\'osi's theory requires a
fundamental length, which they provided in a modified theory.

Ellis, Mohanty \& Nanopoulos (1989) base their theory on a non-unitary
modification of the Hamiltonian evolution equation caused by wormholes
interacting with a microscopic system.

Significant effects which might  distinguish any of these theories from
ordinary quantum mechanics appear only for systems with mass
approaching the  Planck mass, and seem unlikely to be detectable in the
foreseeable future.  In space-time PSD the effects appear for much
smaller systems.

\vfill\eject
{\bf 5. Sharp quantum theories and experiment}

In a letter of 27 May 1926 to Schr\"odinger on his recently
proposed wave equation, H A Lorentz pointed out that a wave
packet, which when moving with the group velocity should represent a
particle, `can never stay together and remain confined to a  small
volume in the long run.  The slightest dispersion in the medium will  pull
it apart in the direction of propagation, and even without that
dispersion it will always spread more and more in the transverse
direction. Because of this unavoidable blurring a wave packet does
not seem to be very suitable for representing things that we want to
ascribe a rather  permanent existence.'

This shows the difficulty of using Schr\"odinger wave packets to
represent classical free particles. The delocalization or dispersion
spreads a wave packet in position and nonlinearities in the dynamics
then spread it in momentum too.  The dispersion is even stronger when
there are particle interactions. And it is exactly this difficulty that
quantum state diffusion, PSD and similar theories overcome, since
quantum state diffusion works in the opposite direction, as shown by
extensive numerical evidence, and theorems (Percival, 1994a). Quantum
state diffusion localizes individual quantum systems into regions of phase
space whose size is determined by Planck's constant.

Bell (1987, p171) uses `sharp formulation of quantum mechanics' to
mean a theory that provides a uniform description of the micro and macro
worlds.  For many decades most physicists believed that there was no
sharp formulation of quantum mechanics, despite the existence of the
sharp pilot wave formulation of de Broglie and Bohm.  Largely due to the
influence of Bell, this is no longer the case. The problem now is that there
are too many different sharp quantum theories.

Apart from the gravitational theories of the previous section, there are
those like the pilot wave theory and many-worlds theories that are in
principle indistinguishable experimentally from ordinary quantum
theory.   There are the theories of consistent or decoherent histories
(Griffiths 1984), for which  the differences are either absent (Omn\'es,
1994) or believed to be so small that they could never be detected by
experiment (Dowker \& Halliwell 1992; Gell-Mann \& Hartle 1993). A
relation between these theories and quantum state diffusion for open
systems is shown by Di\'osi et al (1995).

But for us the important sharp theories are those like PSD that
localize the de Broglie wave packets, and so overcome Lorentz's problem
explicitly.  These theories might be distinguished from ordinary quantum
theory  experimentally. Some experimental tests have been discussed by
Ellis et al, (1984) and Pearle (1984).  However most of the
theories  have free parameters.  For example the
theories of Ghirardi, Rimini and Weber (1986) and Ghirardi, Pearle and
Rimini (1990)  are two-parameter theories.   Recently Pearle and Squires
(1995) have suggested that the gravitational curvature scalar causes the
wave function collapse in these theories.  Those theories which depend on
energy localization (Bedford and Wang 1975, 1977; Gisin 1989;  Milburn
1991),  including PSD1, are one-parameter theories, in which the
parameter determines the rate of energy localization.

In  these theories, the set of  parameters can be chosen with a
wide range of values that are consistent with all experiments to date.
The range can be reduced by further experiment, but in every case there
are possible values which are so extreme that there is no hope of
measuring them.  There is no  accessible critical experiment that is
guaranteed to distinguish between any of these sharp theories and
ordinary quantum theory, which is frustrating for those who might want
to test them experimentally.

The original primary state diffusion theory of PSD1 had one free time
parameter $\tau_0$, which was conjectured to be close to the Planck
time.  That conjecture is here confirmed by changing the foundations of
the theory, but leaving the rest of PSD theory intact, so that a value of
$\tau_0$ close to the Planck time is {\it derived}.  With these new
foundations, PSD becomes a {\it rigid} theory with no free parameters,
except for a factor $C$ of order unity, and with reasonable prospects for
critical experiments using matter interferometry.  It was shown in PSD1
that matter interferometry experiments have already been done
that  were only about a factor of 2000 from distinguishing between
space-time PSD and ordinary quantum mechanics (Kasevich \& Chu 1991),
that is, showing one or the other to be wrong.  Matter interferometry is a
rapidly developing field, so  space-time PSD could be subjected to such a
critical experimental test within the next decade.

This would be a laboratory experiment in quantum gravity.

{\bf Acknowledgements} I thank J Charap, G Ellis, C Hull, C Isham, C Foot
and D Pritchard for helpful communications, T  Brun for introducing me to
the Feynman lectures on gravitation, the atomic theory group at the
University of Freiburg for their stimulating hospitality and the Alexander
von Humboldt Foundation for their generous support.

\vfill\eject
{\bf References}

Bedford, D.  \& Wang, D. 1975 Towards an objective interpretation
of quantum mechanics. {\it Nuo. Cim. B} {\bf 26}, 313-315; 1977 A
criterion for spontaneous state vector reduction. {\it Nuo. Cim. B}
{\bf 37}, 55-62.

Bell, J. S. 1987 {\tenit Speakable and Unspeakable in Quantum
Mechanics}  Cambridge: Cambridge University Press.

Di\'osi, L. 1987 Universal master equation for the gravitational violation
of quantum mechanics. {\it Phys. Lett. A} {\bf 120}, 377-381.

Di\'osi, L. 1989 Models for universal reduction of macroscopic
quantum fluctuations. {\it Phys. Rev. A } {\bf 40}, 1165-1173.

Di\'osi, L. 1992 Quantum measurement and gravity for each
other. In {\it Quantum Chaos, Quantum Measurement; NATO ASI Series
C: Math. Phys. Sci. 357} eds Cvitanovic P., Percival, I. C. \&  Wirzba A.,
Dordrecht: Kluwer, 299-304.

Di\'osi, L., Gisin, N., Halliwell, J. J. \& Percival, I. C. 1995
 Decoherent Histories and Quantum State Diffusion.
{\it Phys. Rev. Lett.} {\bf 74} 203

Di\'osi, L.,  \& Luk'acs, B. 1993a \Ka 's Quantum
space-time generates neutron star density in vacuum. {\it Il Nuo. Cim.}
{\bf 108B} 1419-1421.

Di\'osi, L., \& Luk'acs, B. 1993b Calculation of X-ray signals from \Ka 's
hazy space-time. {\it Phys. Letts. A} {\bf 181} 366-368.

Dowker, H. F. \& Halliwell, J. J. 1992 Quantum mechanics of
history. The decoherence functional in quantum mechanics. {\it Phys.
Rev.}  {\bf D46}, 1580-1609.

Ellis, J., Hagelin, J. S., Nanopoulos, D. V. \& Srednicki, M. 1984
Search for violations of quantum mechanics. {\it Nuc. Phys B} {\bf 241},
381-405.

Ellis, J., Mohanty, S., \& Nanopoulos, D.V.1989 Quantum gravity and the
collapse of the wave function. {\it Phys. Letts. B} {\bf 221}, 113-119.

Feynman, R. P., Moringo, F. B. \& Wagner, W. G. 1962-3 {\it Lectures on
Gravitation} Copyright: California Institute of Technology 1973.

Gardiner, C. W. 1985 {\it Handbook of stochastic methods}
Berlin: Springer.

Gardiner, C. W. 1991 {\it Quantum noise}
Berlin: Springer.

Gell-Mann, M. \& Hartle, J. B. 1993 Classical equations for
quantum systems. {\it Phys. Rev.} {\bf
D47}, 3345-3382.

Ghirardi, G.-C., Grassi, R. \& Rimini, A.. 1990 A continuous spontaneous
reduction model involving gravity {\it Phys. Rev. A} {\bf 42}, 1057-1064.

Ghirardi, G.-C., Rimini, A. \& Weber, T. 1986 {\it Phys. Rev. D}
{\bf 34}, 470-491.

Ghirardi, G.-C., Pearle, P., \& Rimini, A. 1990 {\it Phys. Rev. A} {\bf 42},
78.

Gisin, N. 1989 Stochastic quantum dynamics and relativity. {\it
Helv. Phys. Acta} {\bf 62}, 363-371.

Gisin, N. \& Percival, I. C. 1992 The quantum state diffusion
model  applied to open systems. {\it J. Phys. A} {\bf 25},
5677-5691.

Gisin, N. \& Percival, I. C. 1993a Quantum state diffusion,
localization and quantum dispersion entropy. {\it J. Phys. A}
{\bf 26}, 2233-2244.

Gisin, N. \& Percival, I. C. 1993b The quantum state diffusion
picture of  physical processes.  {\it J. Phys. A} {\bf 26},
2245-2260.

Griffiths, R. 1984 Consistent histories and the interpretation of
quantum mechanics. {\it J. Stat. Phys.} {\bf 36}, 219-272.

Grishchuk, L., Haus, H. A. \& Bergman 1992 Generation of squeezed
radiation from vacuum in the cosmos and the laboratory. {\it Phys. Rev.
D} {\bf 46}, 1440-1449.

Hawking, S. 1982 Unpredictability of quantum gravity.
{\it Commun. Math. Phys.} {\bf 87}, 395-415.

K\'arolyh\'azy, F. 1966 Gravitation and quantum mechanics of
macroscopic objects. {\it Nuo. Cim. A} {\bf 42}, 390-402.

K\'arolyh\'azy, F., Frenkel, A \& Luk\'acs B. 1986 On the possible role of
gravity in the reduction of the wave function. In Penrose \& Isham (1986)

Kasevich, M., \& Chu, S. 1991 Atomic interferometry using stimulated
Raman transitions. {\it Phys. Rev. Letts.} {\bf 67}, 181-184.

Lindblad, G. 1976 On the generators of quantum dynamical
semigroups. {\it Commun. Math. Phys.} {\bf 48}, 119-130.

Milburn, G. J. 1991 Intrinsic decoherence in quantum mechanics
{\it Phys. Rev. A} {\bf 44}, 5401-5406.

Omn\`es, R. 1994 {\it The interpretation of quantum mechanics.}
Princeton University Press.

Pauli, W. 1967 {\it Theory of Relativity} Oxford: Pergamon.

Pearle, P. 1984 Experimental tests of dynamical state vector reduction.
{\it Phys Rev D} {\bf 29}, 235-240.

Pearle, P. \& Squires, E. 1995 Gravitation, energy conservation and
parameter values in collapse models. Preprint, Department of
Theoretical Physics, University of Durham DTP/95/13.

Penrose, R. \& Isham, C. J. (eds) 1986 {\it Quantum Concepts in Space
and Time} Oxford: Clarenden, Oxford Science Publications.

Penrose, R. 1986 Gravity and state vector reduction. In Penrose
\& Isham (1986) 129-146.

Percival, I. C. 1994a Localization of wide open quantum
systems. {\it J. Phys. A} {\bf 27}, 1003-1020.

Percival, I. C. 1994b Primary state diffusion. {\it Proc. Roy. Soc. A} {\bf
447} 189-209. (Reference PSD1)

Weinberg, S. 1972 {\it Gravitation and Cosmology} New York: Wiley.

\end